\documentclass{amsart}
\usepackage{amssymb}
\usepackage[dvips]{graphicx}
\usepackage{amsmath}
\usepackage{dsfont}
\usepackage[ps2pdf,colorlinks=true,pdfpagemode=UseNone, urlcolor=blue,linkcolor=blue,citecolor=red]{hyperref}
\usepackage{url}

\usepackage[english,francais]{babel}

\newtheorem{theorem}{Theorem}[section]
\newtheorem{lemma}[theorem]{Lemma}
\newtheorem{proposition}[theorem]{Proposition}

\newtheorem{e-definition}[theorem]{Definition\rm}

\def\og{\leavevmode\raise.3ex\hbox{$\scriptscriptstyle\langle\!\langle$~}}
\def\fg{\leavevmode\raise.3ex\hbox{~$\!\scriptscriptstyle\,\rangle\!\rangle$}}

\begin{document}

\title{On the Periodic Orbits of the Circular Double Sitnikov Problem}
\author{Hugo Jim\'enez P\'erez},
\email{hjp@fciencias.unam.mx}
\author{Ernesto A. Lacomba}
\email{lace@xanum.uam.mx}

\address
{Departamento de Matem\'aticas, Universidad Aut\'onoma Metropolitana,
Unidad Iztapalapa, C.P. 09340, M\'exico D.F. }

\maketitle
\begin{abstract}
	We introduce a restricted four body problem in a 2+2 configuration
	extending the classical Sitnikov problem to the Double Sitnikov problem.
	The secondary bodies are moving on the same perpendicular line to the 
	plane
	where the primaries evolve, so almost every solution is a collision orbit.
	We extend the solutions beyond collisions with a symplectic 
	regularization and study 
	the set of energy surfaces that contain periodic orbits.

\vskip 0.5\baselineskip
\end{abstract}
\section{Introduction}
We extend the classical Sitnikov problem \cite{Sit1} that is a
special case of the restricted three body problem to a restricted
four body problem in a 2+2 configuration. This configuration consists
in two massive bodies evolving on keplerian orbits around their 
center of masses and two infinitesimal bodies evolving on the perpendicular
line that cross the center of masses.
   The Double Sitnikov Problem consists in determining the evolution of the 
infinitesimal bodies 
under the Newtonian attraction of the massive bodies. 
Since the evolution of the secondaries
is collinear we are interested in solutions with elastic bouncing at collisions. In this note, we consider the 
circular restricted case of the double Sitnikov problem that is the 
integrable case. We study 
the periodic orbits on resonant tori. 

\section{The Circular Double Sitnikov Problem}
We consider that primaries are evolving in circular keplerian orbits and 
each one has mass $m_1=m_2=\frac{1}{2}$. The secondary masses are related 
by $\nu=c\ \mu$ with $c\in]0,1]$. We choose the normalized masses 
$\alpha=\frac{1}{1+c}$ and $\beta=1-\alpha$. With these conditions 
the total energy of the system depends on $\alpha$ and $\mu$ as 
parameters as follows 
\begin{eqnarray}
  H &=& \frac{1}{2} {\bf p}^T  M^{-1} {\bf p} - \frac{\alpha}{\sqrt{q_3^2+1/4}} - \frac{\beta}{\sqrt{q_4^2+1/4}} -\mu\frac{\beta}{q_3-q_4};
  \label{eqn:ham03}
\end{eqnarray}
where
\begin{eqnarray}
M=\left( 
\begin{array}{cc}
  \alpha & 0\\
  0 &\beta 
\end{array}
\right).
\end{eqnarray}
The Hamiltonian vector field is obtained by $i_{X_H}\omega=dH$. The configuration space will be $\mathcal P=\{ q_3>q_4\}$ and the 
flow $\varphi(x,t)=\varphi_t(x)$ is not complete by the singularity due to 
collision. 
To avoid the singularity and to
extend analytically the equations to the hyperplane $q_3=q_4$ we perform a
symplectic regularization with the mapping $\rho:T^*\mathbb R^2\to T^*\mathbb R^2$
defined through the generating function of second type
\begin{eqnarray*}
  W({\bf Q},{\bf p})= p_3\left( Q_4+\beta\frac{Q_1^2}{2}\right) + p_4\left( Q_4-\alpha\frac{Q_3^2}{2} \right) 
\end{eqnarray*}

Then the mapping $\rho: ({\bf Q},{\bf P})\mapsto ({\bf q},{\bf p})$ 
is such that $\rho ^*(\sum_{i} dp_i\wedge dq_i)=\sum_{i} dP_i\wedge dQ_i$
and, therefore $\rho \in Sp(T^*\mathbb R^2)$.
Also, we consider the time rescaling
$ \frac{dt}{d\tau} = \alpha\beta Q_3^2$.
The regularized Hamiltonian function is $L = \alpha\beta Q_3^2
\left( H-h \right)\circ \rho$
this Hamiltonian function depends on $\alpha$, $h$ and $\mu$ as parameters 
and is valid only in the 
energy level $L= 0$ for each $h$ fixed.
If ${\bf z}=(Q_3, Q_4,P_3,P_4)$ we have explicitly
\begin{eqnarray*}
  L &=& \frac{1}{2} \left( \alpha\beta\ P_4^2 Q_3^2 + P_3^2\right) - 2\alpha\beta^2 \mu
   \\
  & & -\alpha\beta Q_3^2\left[ \frac{2\alpha}{\sqrt{\left(2 Q_4+\beta Q_3^2\right)^2+1}}
  + \frac{2\beta}{\sqrt{\left(2 Q_4-\alpha Q_3^2\right)^2+1}} +h \right].
\end{eqnarray*}
Obtaining $\lim_{\mu\to 0} L_h(z;\alpha,\mu)=L_h(z;\alpha,0)$ and reversing 
the process, since $\alpha\beta Q_3^2$ is not identically zero, 
we recover the Hamiltonian function without the term $\mu\frac{\beta}{q_3-q_4}$.
Finally, to extend the solutions in a continuous way beyond collisions 
(with elastic bouncing)
is necessary that the linear momentum be conserved and it is 
possible if and only if $c=1$. We have the following
\begin{proposition}
	In the circular double Sitnikov problem if $\mu=\nu$ then the flow $\varphi_t(x)$ of 
	the limiting case $\mu\to 0$ can be extended to
	a complete flow in a natural way.
\end{proposition}
Then the Hamiltonian system of the double Sitnikov problem is
the triplet $(T^*\mathbb R^2,\omega,H)$ where 
\begin{eqnarray*}
  \omega = \sum_{i=3}^4 dp_i\wedge dq_i,\hspace{10pt}{\rm and}\hspace{10pt}
  H &=& \frac{1}{2} |{\bf p}|^2 - \frac{1}{\sqrt{q_3^2+1/4}} - \frac{1}{\sqrt{q_4^2+1/4}},\hspace{10pt}{\bf p}=(p_3,p_4).
\end{eqnarray*}
\section{Periodic Orbits}
\begin{theorem}
   The action-angle coordinates for the {\rm circular double Sitnikov problem}
   takes the form
  \begin{eqnarray}
	  J(h_i) = \frac{\sqrt{2}}{\pi}\left( 2 E(k_i) - K(k_i) - \Pi(2k_i^2,k_i) \right) \hspace{30pt}
    \theta_i(t;h_i) = \frac{1}{\Omega_i} t(\nu,k) + \theta_{0,i} \label{eqn:Jh1}
  \end{eqnarray}
  where $\Omega_i=\frac{\sqrt{2}}{4\pi(1-2 k_i^2)}(2E(k_i)-K(k_i)+\Pi(2k_i^2,k_i))$
  is the return time of the secondaries, $k_i=\frac{\sqrt{2+h_i}}{2}$ and
   $\theta_{0,i}$,for $i=3,4,$ are constants determined by the initial conditions.
  \label{teo:action}
\end{theorem}

\begin{theorem}
	The solutions for the circular double Sitnikov problem can be written as

\begin{eqnarray*}
	(q_3,p_3,q_4,p_4)(t)= 
	\left( \frac{k_3\ s(\nu_3)\ d(\nu_3)}{1-2 k_3^2\ s^2(\nu_3)},2\sqrt{2}k_3\ c(\nu_3),
	\frac{k_4\ s(\nu_4)\ d(\nu_4)}{1-2 k_4^2\ s^2(\nu_4)},2\sqrt{2}k_4\ c(\nu_4) \right),
\end{eqnarray*}
where $\nu_i$ are functions of $t$ obtained inverting the function
\begin{eqnarray}
	t&=& \int \frac{\sqrt{2}}{4(1-2k^2sn(\nu_i)^2)^2}d\nu_i, \label{eqn:tint}
\end{eqnarray}
and $s(\nu_i)\equiv sn(\nu_i(t),k_i)$,  $c(\nu_i)\equiv cn(\nu_i(t),k_i)$,  
$d(\nu_i)\equiv dn(\nu_i(t),k_i)$ are the sine, cosine, and delta amplitude 
Jacobi elliptic functions, and $k_i=\frac{\sqrt{2+h_i}}{2}$ for $i=3,4$.\label{teo:sol}
\end{theorem}

It is possible to integrate the expression 
(\ref{eqn:tint}) with elliptic functions
and elliptic integrals to obtain  
\begin{eqnarray}
	t&=& \frac{\sqrt{2}}{8(1-2k^2)}\left[2 E(\nu) - \nu +\Pi(\nu,2k^2)
	-4k^2\frac{sn(\nu)cn(\nu)dn(\nu)}{1-2k^2sn(\nu)^2}\right] + C,\hspace{10pt}\label{eqn:timet}
\end{eqnarray}
where $C$ is an arbitrary constant of integration. In \cite{Cor2} the reader will 
find a nice and complete study of this function.

\begin{e-definition} 
We say that $\varphi(t)$ is a periodic solution of period
$\tau$ with $\tau>0$ if $\varphi(t+\tau)=\varphi(t)$ for all $t\in\mathbb R$
and there does not exist $\hat\tau\in (0,\tau)$ such that 
$\varphi(t+\hat\tau)=\varphi(t)$, i.e., $\tau$ is the minimum period.
\end{e-definition}

We will use the notation $(p,q,n)=1$ to mean
that the {\it greatest common divisor} is $gdc(p,q,n)=1$, in other words,
that the three numbers have not common factors at the same time.
\begin{proposition}
      For every periodic solution of the double Sitnikov problem there exist 3-plets
$(p,q,n)\in\mathbb Z^3$ such that $(p,q,n)=1$, and $p>\frac{q}{2\sqrt{2}}$ 
and $p>\frac{n}{2\sqrt{2}}$ holds.
  The periods of these solutions are related to the partial energies by 
$\tau = 2p\pi = q T( h_1) = n T( h_2)$.
\end{proposition}

\begin{e-definition}
\label{def:surface}
	We say that an energy surface $\Sigma_{h_*}=H^{-1}(h_*)$
\emph{accepts} a periodic solution if there exists $(p,q,n)\in \mathbb N^3$  
with the properties: P1) $(p,q,n)=1$, and P2) $p>\frac{q}{2\sqrt{2}}$, $p>\frac{n}{2\sqrt{2}}$ 
such that 
$ h_*=	T^{-1}\left(\frac{2\pi p}{q}\right)+
	T^{-1}\left(\frac{2\pi p}{n}\right) $
We denote the set of the constant energy surfaces that accept periodic
orbits as
\begin{eqnarray*}
	\mathfrak M &=& \left\{ \Sigma_h = H^{-1}(h)| 
 	h=h_*
	\right\}.
\end{eqnarray*}

\end{e-definition}

\begin{theorem}
	In the circular double Sitnikov problem there exists a countable 
	number of energy surfaces $\Sigma\in\mathfrak M$ that 
	contains resonant tori foliated by periodic orbits. Moreover, 
	the set of values $h_*\in H(T^*\mathbb R^2)\subset \mathbb R$ 
	such that $\Sigma_{h_*}
	\in \mathfrak M$
is dense in (-4,0) and have zero measure in $\mathbb R$.\label{teo:tori}
\end{theorem}

The proof of the theorem is an immediate consequence of the two following lemmas

\begin{lemma}\label{prop:exist}
	For each $n\in \mathbb N$ the circular double Sitnikov problem has  
	periodic solutions of period $2n\pi$
\end{lemma}
 We will just exhibit at least one periodic solution of period
$\tau=2N\pi$. This is immediate from the fact that there exists such periodic
solutions in the circular (classical) Sitnikov problem.

{\bf Proof.-} 
For any $N\in\mathbb N$ we can choose the combination
$p=N$ and $q=n=1$ that produce $(p,q)=1$ and  $(p,n)=1$ with
$p>\frac{q}{2\sqrt{2}}$ and $  p>\frac{n}{2\sqrt{2}}$
and the proposition $2.8$ in \cite{Cor2} assures that there exists
$ h_1,h_2\in (-2,0)$ such that 
$T( h_1)=\frac{2\pi p}{q}$  and $ T(h_2)=\frac{2\pi p}{n}$
then the hypersurface $H^{-1}( h_1+ h_2)$ contains a torus foliated
by a family of periodic orbits with period
$\tau = 2\pi N= T( h_1) = T( h_2)$ $\blacktriangle$
\begin{lemma}
	For each $N\in\mathbb N$ fixed,
	the circular double Sitnikov problem
	have a finite number of tori foliated
	by periodic orbits with	period $\tau=2N\pi$. The number
\begin{eqnarray*}
	8 N\varphi(N) +  
	\sum_{q<2\sqrt{2}N}\varphi(q);\hspace{30pt} {\rm with}\hspace{10pt} 
\varphi(p)= p \prod_{n|p}\left( 1-\frac{1}{n} \right),
\end{eqnarray*}
is an upper bound (although is not an optimum bound). $\varphi(p)$ is
the {\it totient} function or {\it Euler's phi function}. 
\end{lemma}
{\bf Proof.-} 
For each $p\in\mathbb N$ fixed there exist $3$-plets $(p,q,n)\in\mathbb N^3$, 
where properties P1 and P2 of Definition \ref{def:surface} holds. Therefore, we search for 
the number $C_p$ of $3$-plets $(p,q,n)=1$ coprimes. It is easy to see that 
for every $q<2\sqrt{2}p$ and $(p,n)=1$, the $3$-plet $(p,q,n)$ does not have
common divisors. This triplets are exactly $(2\sqrt{2}p)\cdot(2\sqrt{2}\varphi(p))=8p\varphi(p)$. 

Additionally, we must add all the couples $(q,n)$ coprime such that $(p,q)$ and $(p,n)$
are not coprime. This means that for each integer $q<2\sqrt{2}p$ with $(p,q)\neq 1$ we must add
the number of coprimes $\varphi(q)$. Then we have 
\begin{eqnarray}
	C_p < 8 p\varphi(p) +  \sum_{q<2\sqrt{2}p}\varphi(q)\label{eqn:Cp} 
\end{eqnarray}
Finally we must eliminate the elements that are in both sets, however the number 
(\ref{eqn:Cp}) is an upper bound of the triplets $(p,q,n)\in\mathbb N^3$ where
properties P1 and P2 holds.

The $3$-plet $(p,q,n)\in\mathbb N^3$ induce a point $x=(2\pi\frac{p}{q},2\pi\frac{p}{n}) 
\in (T(h_3),T(h_4))$ such that the lagrangian torus $\mathbb T =(\mu^{-1}\circ \mathcal 
T^{-1})(x)$ is foliated by periodic orbits of period $2N\pi$, therefore it is a resonant 
torus $\mathbb T_{Res}\subset T^*\mathbb R^2$. 
$\blacktriangle$

{\bf Proof.-(of Theorem \ref{teo:tori})}
The first part of theorem is a consequence of the fact that the countable union 
of finite sets is a countable set. Using Lemma 2 and Lemma 3 we have that the number of 
resonant tori are countable, and since each torus belongs to exactly one energy surface,
the set $\mathfrak M$ is countable too.

We define the map $\mathcal T:\mathfrak g^*\to \mathbb R^2$ by
$(h_3,h_4)\mapsto \left( \frac{T(h_3)}{2\pi},\frac{T(h_4)}{2\pi} \right)$.
For each rational point 
$y\in Img (\mathcal T)$ with $y=(\frac{r}{s},\frac{u}{v})$,
$(r,s)=1$ and $(u,v)=1$, we construct the point $(\frac{ru}{g},\frac{su}{g},\frac{rv}{g})\in
\mathbb N^3$ where $g=gcd(ru,su,rv)$. Since this point fills properties P1 and P2 of Definition
\ref{def:surface}, there exists a resonant torus foliated by periodic orbits with period 
$\tau = 2\frac{ru}{g}\pi = \frac{su}{g} T(h_3)=\frac{rv}{g}T(h_4)$
The set of rational values of $\mathcal T$ defined by $RP:=Img(\mathcal T)\cap \mathbb Q^2$ 
is a dense subset of zero measure in $Img(\mathcal T)$.
The mapping $\mathcal T$ is continuous and then $\mathcal T^{-1}(RP)\subset \mathfrak g^*$
is a dense subset in the image of the momentum map $\mu$. Now we construct the function 
$\mathcal H:\mathfrak g^*\to \mathbb R$ such that sends $x=(h_3,h_4)\mapsto h_3+h_4$. It is 
immediate that $\mathcal H(\mathcal T^{-1}(RP))\subset(-4,0)$ is a dense subset by 
continuity, and have zero measure
since $RP$ is a countable set.
$\blacktriangle$

\section*{Acknowledgements}
 Research partially done during an academic stay of the first
author at the IMCCE institute of the {\it Observatoire
de Paris} and supported by CoNaCyT through Ph.D. fellowship No. 184728.
First author is grateful to the ADS team of the IMCCE Institute for its 
hospitality specially to L. Niederman for the useful discussions and ideas.

\end{document}